\documentclass[letterpaper,10pt]{article} 

\usepackage{opticameet3} %
\usepackage{tikz}
\usepackage{pgfplots}
\tikzset{
  every mark/.append style={solid},
}
\pgfplotsset{compat=1.18}
\usepackage{caption}
\usepackage[keep-defaults, Tol]{colorblind}

\newcommand\authormark[1]{\textsuperscript{#1}}

\usepackage{amsmath,amssymb}
\usepackage[colorlinks=true,bookmarks=false,citecolor=blue,urlcolor=blue,runcolor=black]{hyperref} %

\newcommand{\Xhat}[0]{\hat{X}}

\newcommand{\ticklabelsize}[0]{\footnotesize}

\definecolor{ptb1}{RGB}{187, 187, 187}
\definecolor{ptb2}{RGB}{ 46,  37, 133}
\definecolor{ptb3}{RGB}{ 51, 117,  56}
\definecolor{ptb4}{RGB}{ 93, 168, 153}
\definecolor{ptb5}{RGB}{148, 203, 236}
\definecolor{ptb6}{RGB}{220, 205, 125}

\begin{document}
\title{Reverse Reconciliation with Soft Information\\for Discrete-Modulation CV-QKD at Long Range}
\vspace{-1em}
\author{Marco Origlia,\authormark{1,2,*} Erdem Eray Cil,\authormark{3} Laurent Schmalen,\authormark{3} and Marco Secondini\authormark{1}}
\address{\authormark{1}TeCIP Institute, Sant'Anna School of Advanced Studies, via Giuseppe Moruzzi 1, 56124 Pisa, Italy\\
\authormark{2}IEIIT, National Council of Research, via Girolamo Caruso 16, 56124 Pisa, Italy\\
\authormark{3}Communications Engineering Lab, Karlsruhe Institute of Technology, Hertzstr. 16, 76187 Karlsruhe, Germany}
\begin{abstract}
  We recently introduced a reverse reconciliation scheme with soft information.
  In this paper, we assess its performance at ultra-low SNR, thus proving that such scheme is a versatile solution to the reverse reconciliation problem.
\end{abstract}
\vspace{-.3em}
\section{Introduction}\vspace{-.6em}
Key agreement protocols, such as quantum key distribution (QKD), need reconciliation
procedures to extract a common sequence out of two correlated random sequences available
to Alice and Bob, respectively, who want to establish private communication.
The main reconciliation scheme employed for long range continuous variable QKD based on a Gaussian modulation~\cite{grosshans_continuous_2002} is multidimensional reconciliation (MDR)~\cite{leverrier_multidimensional_2008}.
Recently~\cite{origlia2025soft,origlia_soft-decoding_2025},
we proposed a procedure for reverse reconciliation with soft information (RRS),
in which Bob discretises the channel output and computes a soft metric that is independent of the discretised symbol.
This metric is subsequently sent over the public channel to improve the mutual information
between Alice and Bob.
In~\cite{origlia_soft-decoding_2025}, we proved that, in terms of BER,
RRS closes the performance gap with PAM-4 in an AWGN channel
for high coding rates.
In this work, we demonstrate that this gap can also be closed at ultra-low SNR, by selecting the appropriate bit labelling,
ultimately proving that RRS could replace MDR for long-distance CV-QKD,
while still being capable of operating equally well for short-range CV-QKD.
\vspace{-.6em}
\section{Summary of Reverse Reconciliation with Soft Information}\vspace{-.6em}
The proposed reconciliation scheme works on any modulation format with independent quadratures,
allowing us to restrict the analysis to a real-valued constellation.
In this work, we consider a PAM-4 %
modulation,
with each symbol $X$ drawn by Alice with a uniform probability.
Bob receives a channel output $Y$ with a Gaussian mixture distribution
and discretises it, obtaining $\Xhat$. We assume that the AWGN variance is known, e.g., through
parameter estimation.
He selects the decision thresholds such that $\Xhat$ has uniform probability mass function.
The symbol $\Xhat$ is demapped onto a pair of bits according to a labelling rule (Gray or natural).
After a sufficient number of channel transmissions,
the bit pairs are collected to form a binary frame $\mathbf{B}$, to be used as a raw key, of which
Bob computes and communicates a syndrome over the public channel.
Alice will recover such frame, by means of the log-likelihood ratios (LLRs) $\mathcal{L}$ and the syndrome.
In this case the LLRs solely depend on the transmitted symbol $X$, thus they can be pre-computed.

The aforementioned operations are sufficient to implement reverse reconciliation with hard information only (RRH).
However, to support Alice, Bob applies a transformation function $g(Y)$ to each channel output,
depending on the discretisation region corresponding to $\Xhat$.
The result $N=g(Y)$ is then disclosed and Alice will account for it when computing the LLRs $\mathcal{L}$
for subsequent syndrome-based error correction.
If $g$ is the (possibly complementary) CDF of the channel output given the discretisation region
$F_{Y|\Xhat}(Y)$,
the result $N$ is a continuous random variable with the identical uniform distribution for every discretisation region.
Equivalently, $N$ is independent of $\Xhat$, from which the raw key $\mathbf{B}$ is derived.
Nonetheless, the disclosed metric increases the mutual information between Alice and Bob,
without increasing an eavesdropper's information about
Bob's decisions.

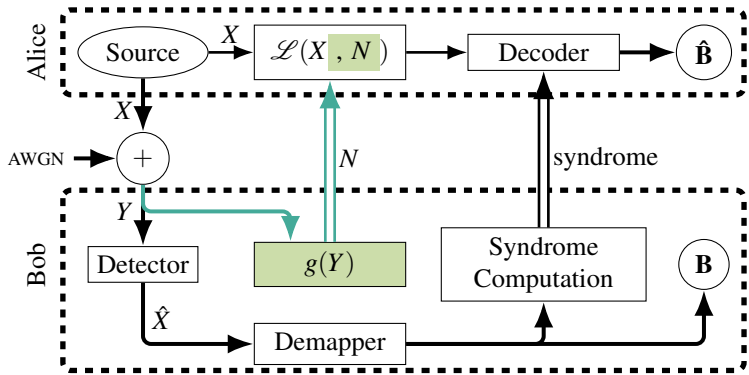
\begin{figure}[!b]
  \vspace{-1em}
  \begin{minipage}[c]{.65\textwidth}
    \begin{tikzpicture}[x=1em, y=1em, baseline=(current bounding box.north)]
  \usetikzlibrary{arrows.meta}
  \usetikzlibrary{shapes}
  \newdimen\yoffset
  \newdimen\ydemapper
  \newdimen\xoffsetreconciliation
  \newdimen\xoffsetsyndrome
  \newdimen\xoffsetB
  \setlength{\yoffset}{4em}
  \setlength{\xoffsetreconciliation}{10em}
  \setlength{\xoffsetsyndrome}{18em}
  \setlength{\xoffsetB}{24em}

  \draw[dashed, line width=2pt, rounded corners=3pt] (0, .6*\yoffset) rectangle (25.5em, 1.4*\yoffset);
  \draw[dashed, line width=2pt, rounded corners=3pt] (0, -1.2em) rectangle (25.5em, -2*\yoffset);

  \node (source) [style={ellipse, draw}] at (3em, \yoffset) {Source};
  \node[style={rectangle, draw}] (detector) at (3em, -\yoffset) {Detector} ;
  \node [style={circle, draw}] (plus_circle) at (3em,0) {$+$};
  \node (Z_text) at (-1em, 0em) {\scriptsize AWGN%
  };
  \node [anchor=east] at (3em, 0.45*\yoffset) {$X$};
  \node [anchor=east] at (3em, -0.5*\yoffset) {$Y$};

  \path [draw, line width=1.5pt, -Latex] (source) to (plus_circle);
  \path [draw, line width=1.5pt, -Latex] (Z_text) to (plus_circle);
  \path [draw, line width=1.5pt, -Latex] (plus_circle) to (detector);

  \node (noisemapper) [draw, style=rectangle, text width=5em, align=center, fill=T-Q-PH3] at (\xoffsetreconciliation, -\yoffset) {$g(Y)$};
  \node (noisedemapper) [draw, style=rectangle, text width=5em, align=center%
  ] at (\xoffsetreconciliation, \yoffset) {$\mathcal{L}(X\colorbox{T-Q-PH3}{, $N$})$}; %
  \path [draw, line width=1pt, double distance=2.5pt, -Latex, T-Q-M3] (noisemapper.north) to (noisedemapper.south);
  \node[anchor=west] at (\xoffsetreconciliation, 0) {$N$};
  \path[draw, line width=1.5pt, rounded corners=3pt, -Latex, T-Q-M3] (plus_circle.south) |- (\xoffsetreconciliation/2, -\yoffset/2) -| (noisemapper.150);

  \node (synenc) [draw, style=rectangle, text width=7em, align=center] at (\xoffsetsyndrome, -\yoffset) {Syndrome Computation};
  \node (syndec) [draw, style=rectangle, text width=5em, align=center] at (\xoffsetsyndrome, \yoffset) {Decoder};
  \path [draw, line width=1pt, double distance=2.5pt, -Latex] (synenc.north) to (syndec.south);
  \node[anchor=west] at (\xoffsetsyndrome, 0) {syndrome};

  \node [style={circle, draw}] (wordestimated) at (\xoffsetB,\yoffset) {$\mathbf{\hat{B}}$};
  \node [style={circle, draw}] (word) at (\xoffsetB,-\yoffset) {$\mathbf{B}$};

  \path [draw, line width=1.5pt, -Latex] (syndec.east) to (wordestimated);
  \path [draw, line width=1pt, -Latex] (source.east) to (noisedemapper.west) ;%
  \path [draw, line width=1pt, -Latex] (noisedemapper.east) %
  to (syndec.west) ;
  \node [anchor=south] at (6.3em, \yoffset) {$X$};

  \node (demapper bob) [style=rectangle, draw, text width=5em, align=center] at (\xoffsetreconciliation, -\yoffset - 3em) {Demapper};
  \path [draw, line width=1.5pt, -Latex, rounded corners=3pt] (detector.south) |- (demapper bob.west);
  \path [draw, line width=1.5pt, -Latex, rounded corners=3pt] (demapper bob.east) -| (synenc.south);
  \path [draw, line width=1.5pt, -Latex, rounded corners=3pt] (demapper bob.east) -| (word.south);
  \node [anchor=west] at (3em, -\yoffset-2em) {$\hat{X}$};

  \node [rotate=90] at (-1em, \yoffset) {Alice};
  \node [rotate=90] at (-1em, -\yoffset) {Bob};
\end{tikzpicture}

  \end{minipage}~\begin{minipage}[c]{.35\textwidth}
    \captionof{figure}{Block diagram of the scheme for reverse reconciliation with soft information (RRS).
      The highlighted elements mark the difference of RRS with respect to the reverse reconciliation with hard information only (RRH).}
  \end{minipage}
  \vspace{-1em}
\end{figure}
\section{Performance at Low SNR}
\vspace{-.6em}

To assess the performance in the low SNR regime,
we estimated the BER of RRS and RRH.
We benchmarked RRS and RRH against PAM-4 in the AWGN channel,
which represents the ultimate limit for RR.
Results were obtained by means of a numerical library
developed by us \cite{origlia_2025_15222665},
and we employed rate-adaptive codes described in \cite{Cil:24}, configured to operate at rates 0.05 and 0.01.

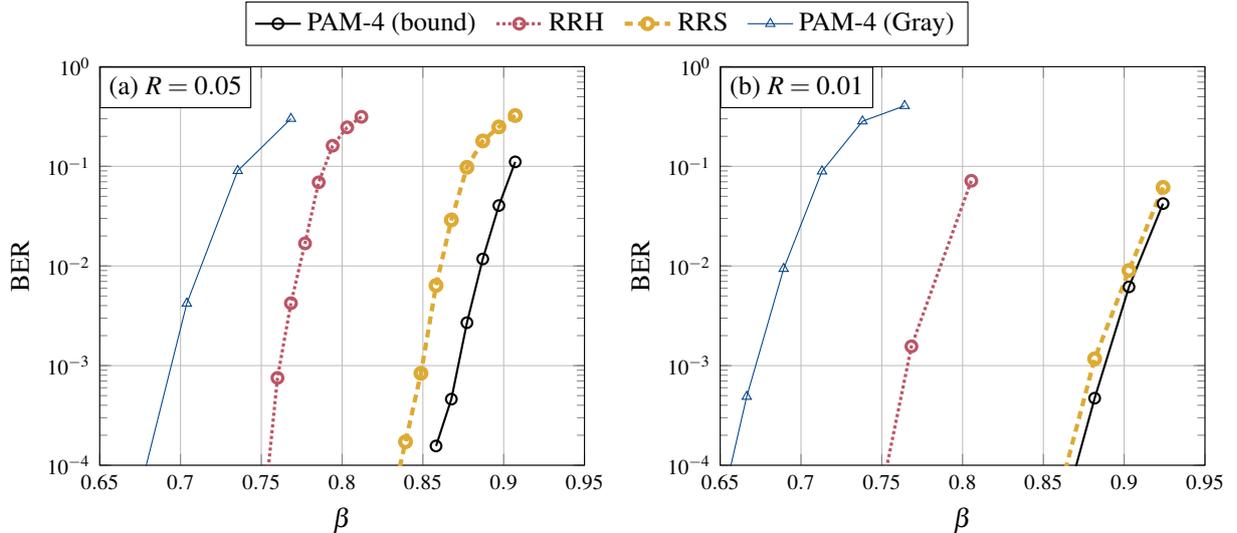
\begin{figure}[!t]
  \centerline{
    \begin{minipage}[b]%
      {\textwidth}
      \begin{center}%
      \ref{legend:pam4-ber-natural-0.05}\end{center}
      \vspace{-1em}
      \begin{tikzpicture}
    \begin{semilogyaxis}[
        xmin=.65,
        ymin=1e-4,
        xmax=.95,
            ymax=1,
            width=0.5\columnwidth,
            legend to name={legend:pam4-ber-natural-0.05},
            legend columns=4,
            every axis legend/.append style={
                /tikz/every even column/.append style={
                    column sep=.5em
                },
                text=black,
            },
            xlabel={$\beta$},
            ylabel={BER},
            every tick label/.append style={
              font=\ticklabelsize,
            },
            grid=major,
            every mark/.append style={solid},
      ]

        \addplot [T-Q-HC0, mark=o, solid, thick] table [
          x expr={ln(2)*0.1/ln(1+10^(\thisrow{SNR}/10))},
          y={BER},
          col sep=comma
        ] {./data_natural_0.05_dir_bps2_snr_-11.000_-10.500_11_it200_sim0250_005000_ferr0250.csv};
        \addlegendentry{PAM-4 (bound)};
        \label{plot:DR}

        \addplot [T-Q-HC3, mark=o, densely dotted, very thick] table [
          x expr={ln(2)*0.1/ln(1+10^(\thisrow{SNR}/10))},
          y={BER},
          col sep=comma
        ] {./data_natural_uniform_0.05_rev-h_bps2_snr_-10.500_-9.500_21_it200_sim0250_005000_ferr0050.csv};
        \addlegendentry{RRH};
        \label{plot:RRH}

        \addplot [T-Q-HC2, mark=o, densely dashed, ultra thick] table [
          x expr={ln(2)*0.1/ln(1+10^(\thisrow{SNR}/10))},
          y={BER},
          col sep=comma
        ] {./data_natural_uniform_0.05_rev-s_bps2_snr_-11.000_-10.500_11_it200_sim0250_010000_ferr0250.csv};
        \addlegendentry{RRS};
        \label{plot:RRS}

        \addplot [T-Q-HC4, mark=triangle, solid] table [
          x expr={ln(2)*0.1/ln(1+10^(\thisrow{SNR}/10))},
          y={BER},
          col sep=comma
        ] {./data_gray_0.05_dir_bps2_snr_-10.750_-8.750_41_it50_sim0250_005000_ferr0250.csv};
        \addlegendentry{PAM-4 (Gray)};
        \label{plot:DRGray}

        \node[anchor=north west, align=right, fill=white, draw] at (axis cs: 0.65,1) {(a) $R=0.05$};
    \end{semilogyaxis}
\end{tikzpicture}
~%
      \begin{tikzpicture}
    \begin{semilogyaxis}[
            xmin=.65,
            ymin=1e-4,
            xmax=.95,
            ymax=1,
            width=0.5\columnwidth,
            legend to name={legend:pam4-ber-natural-0.01},
            legend columns=3,
            every axis legend/.append style={
                /tikz/every even column/.append style={
                    column sep=.5em
                }
            },
            xlabel={$\beta$},
            ylabel={BER},
            every tick label/.append style={
                font=\ticklabelsize
            },
            grid=major
      ]

        \addplot [color=T-Q-HC3, mark=o, mark options={solid}, densely dotted, very thick] table [
          x expr={ln(2)*0.02/ln(1+10^(\thisrow{SNR}/10))},
          y={BER},
          col sep=comma
        ] {./data_natural_0.01_uniform_rev-h_bps2_snr_-17.600_-17.200_3_it200_sim200_2000000_ferr100.csv};

        \addplot [color=T-Q-HC0, mark=o, solid, thick] table [
          x expr={ln(2)*0.02/ln(1+10^(\thisrow{SNR}/10))},
          y={BER},
          col sep=comma
        ] {./data_new_natural_0.01_dir_bps2_snr_-18.200_-17.800_3_it200_sim200_2000000_ferr100.csv};

        \addplot [color=T-Q-HC2, mark=o, mark options={solid}, densely dashed, ultra thick] table [
          x expr={ln(2)*0.02/ln(1+10^(\thisrow{SNR}/10))},
          y={BER},
            col sep=comma
        ] {./data_new_natural_0.01_uniform_rev-s_bps2_snr_-18.200_-17.800_3_it200_sim200_2000000_ferr100.csv};

        \addplot [color=T-Q-HC4, mark=triangle, solid] table [
          x expr={ln(2)*0.02/ln(1+10^(\thisrow{SNR}/10))},
          y={BER},
          col sep=comma
        ] {./data_gray_0.01_dir_bps2_snr_-18.000_-16.500_51_it50_sim0250_010000_ferr0100.csv};
        \node[anchor=north west, align=right, fill=white, draw] at (axis cs: .65,1) {(b) $R=0.01$};
    \end{semilogyaxis}
\end{tikzpicture}

      \vspace{-1.5em}
      \captionof{figure}{Bit error rate characteristics for different reconciliation schemes.}
      \label{fig:ber}
    \end{minipage}
  }
  \vspace{-1em}
\end{figure}

The BER estimation was carried out through 2 sets
of Monte-Carlo simulations, the results of which are shown in Fig.~\ref{fig:ber},
where the SNR values have been converted to reconciliation efficiency $\beta$ against the Gaussian channel.
In each set of simulations, we selected either
code rate 0.05 (Fig.~\ref{fig:ber}~(a)) or 0.01 (Fig.~\ref{fig:ber}~(b)).
First, we simulated a PAM-4 system in an AWGN channel, to be used as a bound for RR,
for both Gray and natural labelling.
In both simulation sets, the natural labelling outperforms the Gray labelling,
unlike communications scenarios where systems are operated at high SNR.
This fact is well understood, as the Gray labelling is not optimal at low SNR \cite{Alvarado2011}.
Then, for subsequent simulations we selected the natural labelling.

At code rate 0.05, the gap between RRH (\ref{plot:RRH}) and the PAM-4 bound (\ref{plot:DR}) is about 0.11 in terms of $\beta$.
However, this gap reduces to about 0.022 when the soft information of RRS (\ref{plot:RRS}) is introduced.
When a code rate of 0.01 is employed instead,
the gap between RRH and the bound is about 0.12, and it reduces to a negligible amount
when RRS is employed, which confirms the improvement of our proposed scheme with respect to
a scheme without the disclosure of the proposed soft metric.

\vspace{-.9em}
\section{Conclusion}
\vspace{-.6em}
In this work, we demonstrate that our RRS scheme outperforms the RRH scheme at ultra-low SNR,
a scenario relevant to long-distance CV-QKD.
Moreover, the BER performance of RRS closely approaches the bound estimated with the performance of PAM-4 modulation in the AWGN channel.
These results prove that RRS is a versatile solution to the reverse reconciliation problem,
which can be adopted both for short- and for long-distance CV-QKD.

\end{document}